\documentclass[12pt]{JHEP}
\usepackage{latexsym,kec}
\title{Supersymmetry without Grassmann Variables}
\author{Kevin Cahill \\ 
New Mexico Center for Particle Physics \\
Department of Physics and Astronomy \\
University of New Mexico \\
Albuquerque, New Mexico 87131-1156 \\
\email{cahill@unm.edu}}
\abstract{Supersymmetry transformations
may be represented by unitary operators
in a formulation of supersymmetry without
numbers that anti-commute.
The physical relevance of this formulation
hinges on whether or not one may add states
of even and odd fermion number,
a question which soon may be settled by experiment.}
\preprint{NMCPP/01-3}
\begin{document}

\setcounter{equation}{0}
\pagestyle{plain}

\section{Introduction}
Supersymmetry usually is formulated 
in terms of anti-commuting numbers 
called Grassmann variables. 
In these formulations supersymmetry 
transformations \(T(\xi)\) are represented by exponentials 
\(T(\xi) = \exp( \xi Q + \bar Q \bar \xi)\)
of the supercharges \(Q_a\)
multiplied by the components \(\xi^a\)
of  a Grassmann spinor~\cite{BW}.
But since the Hilbert space \(\cal{H}\) of quantum mechanics
is a vector space over the complex numbers,
and not over a field containing Grassmann variables,
the image \(T(\xi)|\psi\rangle\) 
of any state \(|\psi\rangle\) not annihilated by
the fermionic generators \(Q_a\) will lie outside of \(\cal{H}\)\@.
One may include states of the form \(T(\xi)|\psi\rangle\)
either by enlarging the Hilbert space \(\cal{H}\) 
or by expressing the transformations of supersymmetry
as unitary transformations on \(\cal{H}\)\@.
The latter course is the subject of this article.
\par
Inasmuch as supersymmetric quantum field theories
are merely ordinary quantum field theories 
with particular fields and parameters,
it is clear that one can 
discuss supersymmetry without the use of
Grassmann variables.  Moreover if \(z^a\) 
is a complex spinor, then the operator
\(G(z) = z Q + \bar Q \bar z\) is hermitian,
\(G(z)^\dag = G(z)\), 
and the operator \(U(z) = \exp(-iG(z))\) is unitary,
\(U(z)^\dag = U(z)^{-1}\)\@. 
But the image \(U(z)|\psi\rangle\) 
of any state \(|\psi\rangle\) not annihilated by
the fermionic generators \(Q_a\) must be
a superposition of states of even and odd fermion number.
It is generally believed, however,
that such states are unphysical because under a 
rotation of \(2\pi\) around any axis, they
change by more than an overall phase~\cite{WWW}.
If this superselection rule is true, then the state
\(U(z)|\psi\rangle\) lies outside of the
physical Hilbert space \(\cal{H}\),
and we gain very little by considering the 
unitary operator \(U(z)\)\@.
\par
But this superselection rule is true or false
according to whether the Lorentz group of nature
is \(SL(2,C)/Z_2\) or \(SL(2,C)\),
and its validity can be settled 
only by experiment because
the consequences of \(SL(2,C)\)
are the same as those of \(SL(2,C)/Z_2\)
except for the absence of superselection rules~\cite{SW}.
Andreev~\cite{AA} has suggested how
this rule may be tested by experiments on mesoscopic 
metal particles at low temperatures.
Efforts to perform such experiments are in progress.
\par
It may make sense therefore to resolve 
some of the technical problems that arise when
one tries to represent the transformations
of supersymmetry by unitary operators. 
The case of the chiral multiplet will
be considered in this paper.

\section{Unitary Supersymmetry Transformations}
In what follows it will be shown how to implement 
supersymmetry transformations on the chiral multiplet 
by means of unitary operators without Grassmann variables.
The supersymmetry transformations 
generated by the operator 
\({\cal G}(\x) = \x Q + \bar Q \bar \x \),
in which \(\x^a\) is a Grassmann variable
and \(Q_a\) a supercharge,
will be recalled, and the transformations 
induced by the generator
\(G(z) = z Q + \bar Q \bar z \),
where \(z\) is a complex number, 
will be discussed.  It will be shown that these
transformations are a symmetry of the theory
because the change in the action density is a divergence
of a current.  The Noether procedure will
then be used to show that the fermionic generators
of this symmetry are the usual supercharges \(Q_a\),
which satisfy the algebra of supersymmetry.
Thus the unitary operator \(U(z) = \exp[-iG(z)]\) 
represents a supersymmetry transformation.
\par
The action density for the chiral multiplet
with superpotential \(W(A)\) is
\beq
{ \cal L} = \ihalf \p_n \bar \psi \bar \s^n \psi
- \ihalf \bar \psi \bar \s^n \partial_n \psi
- \partial_n \bar A \partial^n A
- | W' |^2 - \half W'' \psi \psi - \half \bar W'' \bar \psi \bar \psi 
\label {Lcn}
\eeq
in which \( W' = \partial W(A)/ \partial A\)\@.
The supercharges \(Q_a\) are
\beq
Q_a  = \sqrt{2} \int \! d^3x \, \left(
\s^m_{a\db} \bar \s^{ 0 \db c }
\psi_c \partial_m \bar A
- i \s^0_{a\db} \bar \psi^\db \bar W' \right) . \label {QR}
\eeq
In a mixed notation they are
\bea
Q_a & = & - \sqrt{2} \int \! d^3x \, \left(
\s^m_{a\db} \psi_b \partial_m \bar A
- i \bar \psi^\da \bar W' \right)
\label {Qup} \\
& = & - \sqrt{2} \int \! d^3x \, \left(
\s^m_{a\db} \psi_b \partial_m \bar A
- i \e^{\da \db} \bar \psi_\db \bar W' \right) 
\label {Qdown}
\eea
or more simply
\beq
Q = \sqrt{2} \int \! d^3x \, \left[ \, \left(
\p_0 \bar A - \vec \s \cdot \nabla \bar A \right) \psi
- \bar W' \, \s^2 \bar \psi \, \right] .
\label {QRss}
\eeq
\par
In the usual formalism with Grassmann variables \(\x^a\),
the ``bosonic'' operator
\beq
\mathcal{G}( \x ) = \x Q + \bar Q \bx 
\eeq
generates in the fields $ A(x) $ and \(\psi(x)\) the changes
\bea
\d_\x A( x ) \equiv \left[ i \mathcal{G}( \x ), A( x ) \right] & = &
i \sqrt{2} \int \! d^3y \, 
\x^a \s^m_{a\db} \psi_b( y ) 
\left[ A( x ) , \partial_m \bar A( y ) \right] \nn\\
& = & \mbox{} - \sqrt{2} \int \! d^3y \,
\x^a \s^0_{a\db} \psi_b( y )
\d( \vec x - \vec y ) \nn\\
& = & \mbox{} \sqrt{2} \x^a \psi_a( x ) = \sqrt{2} \x \psi ,
\label {dA}
\eea
and
\bea
\d_\x \psi_a(x) & \equiv & \left[ i \mathcal{G}( \x ), \psi_a(x) \right] \nn\\
& = & \mbox{} \sqrt{2} \! \int \! \! d^3y 
\left[ \psi_a(x), \x^b \e^{\db \dc} \bar \psi_\dc(y) \bar W'
+ i \bar \psi_\dc(y) \s^m_{c \db} \bx^\db \partial_m A(y) \right] 
\nn\\
& = & \mbox{} \sqrt{2} \left(
- \x^b \e^{\db \da} \bar W'
+ i \s^m_{a \db} \bx^\db \partial_m A(x) \right)
\nn\\
& = & \mbox{}
\sqrt{2} \left( \x^b \e_{b a} \bar W'
+ i \s^m_{a \db} \bx^\db \partial_m A(x) \right)
\nn\\
& = & \mbox{} \sqrt{2} i \s^m_{a \db} \bx^\db \partial_m A(x)
- \sqrt{2} \x_a \bar W',
\label {dpsi}
\eea
which is the usual result 
if we use $ - \bar W' = F $,
where \(F\) is the auxiliary field. 
\par
Exponentials of the generator
\( \mathcal{G}( \x ) = \x Q + \bar Q \bx \)
are not unitary operators
because they involve Grassmann variables.
Can one avoid these anti-commuting variables?
Let us consider using generators
\( G( z ) \) that are complex linear
forms in the supercharges \( Q \) and \( \bar Q \)
\beq
G( z ) = z Q + \bar Q \bar z
\label {G(z)}
\eeq
where \( z^a \) is a complex spinor. 
Now the change in the field $ A(x) $ is
\bea
dA( x ) \equiv \left[ \, i G( z ) , A( x ) \, \right] & = &
\mbox{} - i \sqrt{2} \int \! d^3y \,
z^a \s^m_{a\db} \psi_b( y )
\left[ \, \partial_m \bar A( y ) , A( x ) \, \right] \nn\\
& = & \mbox{} - \sqrt{2} \int \! d^3y \,
z^a \s^0_{a\db} \psi_b( y )
\d( \vec x - \vec y ) \nn\\
& = & \mbox{} \sqrt{2} z^a \psi_a( x ) = \sqrt{2} z^a \psi_a ,
\label {dzA}
\eea
which is the same as (\ref {dA})
except that the Grassmann spinor \( \x \) 
has been replaced by the complex spinor \( z \).
The conjugate change is
\beq
d\bar A = \sqrt{2} \bar \psi_\da \bar z^\da .
\label {dzbarA}
\eeq
\par
This procedure will not work, however,
for the Fermi field \( \psi \).
Instead we must write \( d \psi \) as an anti-commutator.
There are several ways of doing this,
but if we want $ d \psi $ to be the adjoint
of $ d \bar \psi $, then we can not have
a single rule for the
change in the product of two spinor fields
irrespective of whether they transform like
$ \psi $ or like $ \bar \chi $.  
We choose to have $ d \psi $ be the adjoint
of $ d \bar \psi $, and so we shall have four
different rules for the change in the
product of two spinor fields.  We define
\bea
d\psi_a(x) & \equiv & - \left\{ i G( z ) , \psi_a(x) \right\} \nn\\
& = & \mbox{} \sqrt{2} \! \int \! \! d^3y
\left\{ z^b \e^{\db \dc} \bar \psi_\dc(y) \bar W'
+ i \bar \psi_\dc(y) \s^m_{c \db} \bz^\db \partial_m A(y) ,
\psi_a(x) \right\}
\nn\\
& = & \mbox{} \sqrt{2} \left(
 z^b \e^{\db \da} \bar W'
+ i \s^m_{a \db} \bz^\db \partial_m A(x) \right)
\nn\\
& = & \mbox{}
\sqrt{2} \left( - z^b \e_{b a} \bar W'
+ i \s^m_{a \db} \bz^\db \partial_m A(x) \right)
\nn\\
& = & \mbox{} \sqrt{2} i \s^m_{a \db} \bz^\db \partial_m A(x)
+ \sqrt{2} z_a \bar W' .
\label{dzpsi}
\eea
The change in the conjugate $\bar \psi $ 
is the conjugate of the change in $ \psi $
\bea
d\bar \psi_\da & \equiv & \left\{ i G( z ) , \bar \psi_\da \right\}  
= \left( - \left\{ i G( z ) , \psi_a(x) \right\} \right)^\dgr 
= ( d \psi_a )^\dgr \nn\\
& = & \mbox{} - \sqrt{2} i z^b \s^m_{b \da} \p_m \bar A 
+ \sqrt{2} \bar z_\da W' .
\label{dzbarpsi}
\eea
Although these formulas differ from expression (\ref {dpsi})
for $ \d\psi $ and its conjugate for $ \d\bar\psi $
by the signs of their second terms, and of course
by the replacement of a Grassmann spinor $ \x $ by 
a complex one $ z $,
we shall see that these sign differences
are appropriate and that supersymmetry can be implemented
by unitary transformations acting on the
states and physical operators of the theory.
\par
The key point is that the physical operators of the theory
contain even powers of the Fermi fields.
Thus the change in the generic product
$ \psi \chi $ of two Fermi fields is
\bea
d(\psi \chi) & = & 
\left [ \, i G ( z ) , \psi \chi \, \right ] \nn\\ 
& = & \mbox{} i G ( z ) \, \psi \chi - \psi \chi \, i G ( z ) \nn\\
& = & \mbox{} i G ( z ) \, \psi \chi + \psi \, i G (z ) \, \chi
- \psi \, i G (z ) \, \chi - \psi \chi \, i G (z ) \nn\\
& = & \mbox{} 
\left \{ i G (z ) , \psi \right \} \chi 
- \psi \left \{ i G (z ) , \chi \right \}
\nn\\
& = & \mbox{} - d \psi \, \chi + \psi \, d \chi ,
\label {dpsiphi}
\eea
in which the spinor indices, which may be different, are suppressed. 
It is easy to see that the other three rules are:
\bea
d( \bar \psi \bar \chi ) = 
\left [ \, i G(z) , \bar \psi \bar \chi \, \right ] 
& = & d \bar \psi \, \bar \chi
- \bar \psi \, d \bar \chi 
\label {barpsibarphi} \\
d( \bar \psi \chi ) = 
\left [ \, i G(z) , \bar \psi \chi \, \right ] 
& = & d \bar \psi \, \chi 
+ \bar \psi \, d \chi 
\label {barpsiphi} \\
d( \psi \bar \chi ) = 
\left [ \, i G(z) , \psi \bar \chi \, \right ] 
& = & - d \psi \, \bar \chi
- \psi \, d  \bar \chi .
\label {psibarchi}
\eea
\par
Let us now consider the effect of these
transformations on the chiral action density (\ref{Lcn}).
The change in $ { \cal L} $ due to the 
changes $ d A $ and $ d \psi $ and their conjugates is
\bea
\lefteqn{
d{ \cal L } = \left [ \, i G(z) , { \cal L } \, \right] =
\ihalf \p_n d \bar \psi \bar \s^n \psi
+  \ihalf \p_n \bar \psi \bar \s^n d \psi
- \ihalf d \bar \psi \bar \s^n \partial_n \psi
- \ihalf \bar \psi \bar \s^n \partial_n d \psi } \hspace {0.9in} \nn \\
& & \mbox{} - \partial_n d \bar A \partial^n A
- \partial_n \bar A \partial^n d A 
- \bar W' W'' d A - W' \bar W'' d \bar A \nn\\
& & \mbox{} - \half W''' d A \psi \psi
- \half \bar W''' d \bar A \bar \psi \bar \psi 
+ \half W'' d \psi \psi - \half W'' \psi d \psi \nn\\
& & \mbox{} - \half \bar W'' d \bar \psi  \bar \psi
+ \half \bar W'' \bar \psi d \bar \psi .
\label {dLcn}
\eea 
The part of $ d{ \cal L} $ that depends upon $ z $ is
\bea
d_z { \cal L} & = &
\ihalf \p_n ( - \sqrt{2} i z \s^m \p_m \bar A ) \bar \s^n \psi
+  \ihalf \p_n \bar \psi \bar \s^n ( \sqrt{2} z \bar W' ) \nn\\
& & \mbox{}
- \ihalf ( - \sqrt{2} i z \s^m \p_m \bar A ) \bar \s^n \partial_n \psi
- \ihalf \bar \psi \bar \s^n \partial_n ( \sqrt{2} z \bar W' ) \nn\\
& & \mbox{} - \partial_n \bar A \partial^n ( \sqrt{2} z \psi )
- \bar W' W'' ( \sqrt{2} z \psi ) \nn\\
& & \mbox{} - \half W''' ( \sqrt{2} z \psi ) \psi \psi
+ \half W'' ( \sqrt{2} z \bar W' ) \psi 
- \half W'' \psi ( \sqrt{2} z \bar W' ) \nn\\
& & \mbox{} - \half \bar W'' ( - \sqrt{2} i z \s^m \p_m \bar A ) \bar \psi
+ \half \bar W'' \bar \psi ( - \sqrt{2} i z \s^m \p_m \bar A ) .
\label {dLcnd}
\eea
Because Fermi fields at equal times anti-commute,
the term proportional to $ W''' $ vanishes.
The two terms proportional 
to $ \bar W' \, W'' $ cancel.
The last two terms may be written as
\beq
\irhalf \bar \psi^\da \s^m_{b\da} z^b \p_m \bar W' = 
\irhalf \bar \psi_\dc \, \e^{\da\dc} \s^m_{b\da} \e^{bd} z_d \, \p_m \bar W'
= \irhalf \bar \psi \bs^m z \, \p_m \bar W'
\eeq
and as
\beq
- \irhalf \bar \psi_\da z^b \s^m_{b\dc} \, \e^{\da\dc} \p_m \bar W' =
- \irhalf \bar \psi_\da \e^{\da\dc} \e^{bd} \s^m_{b\dc} z_d \, \p_m \bar W'
= \irhalf \bar \psi \bs^m z \, \p_m \bar W' .
\eeq
So the change $ d_z { \cal L} $ in the action density is 
\bea
d_z { \cal L} & = &
{1 \over \sqrt{2} } z \s^m \bs^n \psi \p_n \p_m \bar A
- {1 \over \sqrt{2} } z \s^m \bs^n  \p_n \psi \p_m \bar A \nn\\
& & \mbox{} - \sqrt{2} z \p^n \psi \p_n \bar A
+ \irhalf \p_n \left( \bar \psi \bs^n z \bar W' \right) .
\label {daL}
\eea
We may write this change $ d_z { \cal L} $ as the total
divergence
\beq
d_z { \cal L} = \p_n K_z^n
\eeq
of the current
\beq
K_z^n = - { 1 \over \sqrt{2} } z \s^m \bs^n \p_m \bar A
- \sqrt{2} z \psi \p^n \bar A 
+ \irhalf \bar \psi \bs^n z \bar W' ,
\label {Kz}
\eeq
which shows that the action is invariant 
under the unitary transformation
\beq
U(z) = e^{-iG(z)}
\label {U}
\eeq
at least for infinitesimal values of the
complex spinor $ z $. 
\par
The Noether current associated with the
susy transformation (\ref{dzA}--\ref{dzbarpsi}) 
of the action density (\ref{Lcn}) is
\begin{eqnarray}
J^n &=& \frac{ i }{ 2 } d \bar \psi \bar \s^n \psi
- \frac{ i }{ 2 } \bar \psi \bar \s^n d \psi
- d \bar A \partial^n A - \partial^n \bar A d A \nn\\
&=& \mbox{} \frac{ i }{ 2 } 
\left( - i \sqr2 z \s^m \partial_m \bar A
+ \sqr2 \bar z W' \right) \bar \s^n \psi
- \frac{ i }{ 2 } \bar \psi \bar \s^n 
\left( i \sqr2 \s^m \bar z \partial_m A + \sqr2 z \bar W' \right) \nn\\
& & \mbox{} - \sqr2 \bar z \bar \psi \partial^n A
- \partial^n \bar A \sqr2 z \psi .
\eea
The part depending on $ z $ is
\begin{equation}
J^n_z = 
\rhalf z \s^m \bar \s^n \psi \partial_m \bar A
- \frac{ i }{ \sqr2 } \bar \psi \bar \s^n z \bar W'
- \sqr2 z \psi \partial^n \bar A .
\eeq
\par
The Noether current $ J^n $  satisfies
\begin{equation}
d {\cal L} = \partial_n J^n ,
\eeq
and so the difference $ J^n - K^n $ of the two currents 
\begin{equation}
S^n = J^n - K^n
\eeq
is conserved
\begin{equation}
\partial_n S^n = 0 .
\eeq
The current $ S^n $ is the supercurrent
of the Lagrange density $ {\cal L} $.
The part $ S^n_z $ that depends upon $ z $ is simply
\begin{equation}
S^n_z = \sqr2 z \s^m \bar \s^n \psi \partial_m \bar A 
- i \sqr2 \bar \psi \bar \s^n z \bar W'.
\eeq
Thus the quantity $ z Q $ is
\begin{equation}
z Q = \int \! d^3x \, S^0_z 
= \int \! d^3x \, \sqr2 z \s^m \bar \s^0 \psi \partial_m \bar A 
- i \sqr2 \bar \psi \bar \s^n z \bar W',
\eeq
and so by the identity
$ \bar \psi \bar \s^0 x = z \s^0 \bar \psi $, 
the supercharges $ Q_a $ are
\bea
Q_a & = & \sqrt{2} \int \! d^3x \, 
\left( \s^m_{a\db} \bar \s^{ 0 \db c } 
\psi_c \partial_m \bar A 
- i \s^0_{a\db} \bar \psi^\db \bar W' \right) .
\label {Q}
\eea
These supercharges, which generate 
the unitary supersymmetry transformations (\ref{U})
and (\ref{dzA}--\ref{dzbarpsi}), 
are the same as the those (\ref{QRss}) that generate 
the Grassmann supersymmetry transformations (\ref{dA}--\ref{dpsi})\@.
\par
We have seen that the action density (\ref{Lcn}) 
is invariant under the unitary transformation (\ref{U})
which is an exponential of an imaginary linear form
in the supercharges (\ref{QR}), that the induced 
complex supersymmetry transformations (\ref{dzA}--\ref{dzbarpsi})
differ somewhat from the usual 
supersymmetry transformations (\ref{dA}--\ref{dpsi}),
and that the supercharges derived by the Noether
technique (\ref{dLcn}--\ref{Q})
from the complex supersymmetry transformations 
(\ref{dzA}--\ref{dzbarpsi}) are the same as the 
conventional supercharges (\ref{QR})\@.
\par
It is straightforward to generalize this
argument to the general chiral theory 
consisting of \(N\) multiplets \(A_i, \psi_i, F_i\)
interacting through an arbitrary analytic
superpotential \(W(A_1,\dots,A_N)\)\@.
In fact when appropriately generalized,
the rules (\ref{G(z)}--\ref{psibarchi})
should apply to any supersymmetric field theory.
Unitary operators without Grassmann variables
can implement supersymmetry transformations
upon the action and upon other
operators that involve even powers of Fermi fields.
\acknowledgments
I should like to thank A.~F.\ Andreev 
and M.~Serna 
for interesting conversations.

\end{document}